\documentclass[prb,twocolumn,superscriptaddress,floatfix]{revtex4-1}
\usepackage{bigints}
\usepackage{yfonts}
\usepackage{hyperref}
\usepackage{graphicx}
\usepackage{amsmath}
\usepackage[dvips]{epsfig}

\begin{document}

\title{Self-sustained deformable rotating  liquid He cylinders: The pure normal fluid $^3$He and superfluid $^4$He cases}

\author{Mart\'{\i} Pi}
\affiliation{Departament FQA, Facultat de F\'{\i}sica,
Universitat de Barcelona. Diagonal 645,
08028 Barcelona, Spain}
\affiliation{Institute of Nanoscience and Nanotechnology (IN2UB),
Universitat de Barcelona, Barcelona, Spain.}

\author{Francesco Ancilotto}
\affiliation{Dipartimento di Fisica e Astronomia ``Galileo Galilei''
and CNISM, Universit\`a di Padova, via Marzolo 8, 35122 Padova, Italy}
\affiliation{ CNR-IOM Democritos, via Bonomea, 265 - 34136 Trieste, Italy }

\author{Manuel Barranco}

\affiliation{Departament FQA, Facultat de F\'{\i}sica,
Universitat de Barcelona. Diagonal 645,
08028 Barcelona, Spain}
\affiliation{Institute of Nanoscience and Nanotechnology (IN2UB),
Universitat de Barcelona, Barcelona, Spain.}

\author{Samuel L. Butler}
\affiliation{Department of Geological Sciences, University of Saskatchewan, Saskatoon, Saskatchewan S7K 2Y6, Canada.}

\author{Jos\'e Mar\'{\i}a Escart\'{\i}n}
\affiliation{Catalan Institute of Nanoscience and Nanotechnology (ICN2),
CSIC and BIST, Campus UAB,
Bellaterra,
08193 Barcelona, Spain.}

\begin{abstract}
 We have studied self-sustained, deformable, rotating liquid He cylinders of infinite length. In the normal fluid $^3$He case, we
 have employed a classical model where only surface tension and centrifugal forces are taken into account, as well as the Density Functional
 Theory (DFT) approach in conjunction with a semi-classical Thomas-Fermi approximation for the kinetic energy. In both approaches,
 if the angular velocity is sufficiently large, it is energetically favorable for the $^3$He cylinder to undergo a shape transition, acquiring  an
elliptic-like cross section which eventually becomes two-lobed.
In the $^4$He case, we have employed a DFT approach that
takes into account its superfluid character, limiting the description to vortex-free
configurations where angular momentum is exclusively stored in capillary waves on a deformed cross section cylinder.
The calculations allow us to carry out a comparison between the rotational behavior of a normal, rotational fluid ($^3$He) and a superfluid,
irrotational fluid ($^4$He).

\end{abstract}
\date{\today}

\maketitle

\section{Introduction}

Helium is the only element in nature that may remain liquid at temperatures close to  absolute zero.
At very low temperatures, liquid helium is the ultimate quantum fluid, able to form nanoscopic droplets and  macroscopic
samples as well. In particular, $^4$He droplets are considered
as ideal matrices for spectroscopic studies of embedded atoms and molecules,
and for the study of superfluidity at the atomic scale, including the study of quantum vortices.

Determining the experimental size and shape of helium droplets is a challenging problem.
First determinations
focussed on droplets made of up to several  thousand atoms produced by the adiabatic
expansion of a cold helium gas.\cite{Toe04} The experiments analyzed the scattering cross section of   Kr atoms  dispersed
by a jet of  $^4$He or  $^3$He droplets using DFT density profiles as input.\cite{Har98,Har01}
 More recently, large helium droplets made  of $10^8-10^{11}$ atoms have been  created by the hydrodynamic instability of a very
 low temperature $(T)$ liquid helium
jet passing through the nozzle of a molecular beam apparatus, as reviewed in Ref. \onlinecite{Tan18}. These large drops
have been analyzed to determine their shape and, in the case of $^4$He, whether they host arrays of quantized vortices or
not.\cite{Gom14}

Two experimental techniques have been used to characterize large helium drops.
Coherent diffractive imaging of x-rays from a free electron laser\cite{Gom14,Oco20,Ver20,Fei21}  gives access to a model-independent determination
of the two-dimensional (2D) projection of the drop density on a plane perpendicular to the x-ray incident direction via iterative phase retrieval
algorithms. Irradiation of helium droplets with intense extreme ultraviolet pulses\cite{Rup17,Lan18}
and subsequent measurement of wide-angle diffraction patterns
provides access to full three-dimensional information. However,
so far the analysis of the densities is model-dependent as their shape
has to be  parameterized using some guessed solid figure which is then used to produce a diffraction pattern which may be compared
to the experimental one.
The conclusion drawn from these experiments, which are mostly on $^4$He drops, is that helium drops are mainly spherical and that only a small fraction of them
(about 7\%)\cite{Lan18} are deformed and host some angular momentum, which is likely
acquired during their passage through the nozzle of the experimental apparatus.
The experimental results have been  compared to calculations made for incompressible viscous droplets only subject
to surface tension  and centrifugal  forces,\cite{Bro80,Hei06,But11,Bal15,Ber17} or based on a Density Functional Theory (DFT) approach\cite{Anc15,Anc18,Pi21}
specifically designed to describe liquid helium.\cite{Dal95,Bar97,Bar06,Anc17}

It has been found\cite{Lan18,Oco20} that spinning $^4$He droplets follow the sequence of shapes characteristic of rotating viscous
droplets.\cite{Bro80,Bal15}
This unexpected result is due to the presence of vortex arrays in the spinning droplets\cite{Anc18,Pi21}
that confer to them  the appearance of  rotating rigid-bodies.

Large $^3$He drops have been detected as well in the coherent diffractive imaging of x-rays experiments.\cite{Ver20,Fei21}
It is worth mentioning that classical and DFT  calculations for $^3$He droplets have yielded very similar
relationships between angular velocity and
angular momentum,\cite{Pi20} which is due to the fact that at the experimental temperatures
($T \sim 0.15$ K)\cite{Sar12} liquid $^3$He behaves as a normal fluid with a finite viscosity.

So far, the most reliable approach to study spinning helium droplets is  the DFT approach. It has, however,  the limitation
that the complexity of DFT  calculations dramatically grows with droplet size, making them prohibitively costly
even for a few tens of thousands  of atoms, which is significantly below typical
experimental sizes. Addressing large droplets
is especially needed to study large vortex array configurations
in $^4$He droplets as well as the recently produced spinning mixed $^3$He-$^4$He
droplets,\cite{Oco21} for which  classical\cite{But20,But22} and DFT\cite{Pib20} calculations are already available.
To circumvent this limitation, it is often resorted to a simpler cylindrical geometry which restricts the calculations
to less demanding 2D configurations
while the basic physics may still be caught by the model. Indeed, self-sustained $^4$He circular and deformed cylinder configurations have been
used to describe the density of spinning $^4$He droplets on the plane of symmetry perpendicular to the rotation axis.\cite{Anc14,Oco20}
We stress that these are self-sustained configurations, and not rotating cylindrical vessels (circular or elliptic) filled with helium for which there exists a
vast literature, see e.g. Refs. \onlinecite{Fet74,Cam79,Don91} and references therein.

In this work we describe self-sustained liquid He cylinders of infinite length under rotation.
The equilibrium and stability of a rotating column of a viscid fluid subject to planar disturbances have been addressed
in detail\cite{Ben91,But22} applying techniques similar to those used to describe rotating viscid droplets.\cite{Bro80,But11} As in the present study,
only translationally symmetric (planar) disturbances leading to non-circular cylinder cross sections have been considered there.
Axisymmetric Rayleigh instabilities, always present in fluid columns, were set aside.

In the case of $^3$He, we have employed a classical model for viscous liquids subject to centrifugal and surface tension forces,\cite{But11} and a normal liquid
DFT\cite{Str87,Bar97} plus semiclassical approach, treating the $^3$He cylinders in the DFT plus rotating Thomas-Fermi (TF)
framework.\cite{Pi20,Gra78} This semiclassical approach is justified by the large number of atoms per unit length in the cylinder.
The DFT-TF method represents a realistic framework allowing  to make  the calculations affordable. It
can be extended  to mixed helium  systems as well.\cite{Bar06,Pib20}
As for droplets,\cite{Mat13a} thermal effects  on the energetics and morphology of the cylinder are expected to be negligible at the experimental
temperatures, so we shall use a  zero temperature method.\cite{Pi20} Zero temperature means here  a very low $T$, but above
$T \sim 2.7$ mK at which  $^3$He becomes superfluid.

In the $^4$He case, we have employed a DFT approach which takes into account its superfluid character,\cite{Anc17} limiting the description to vortex-free
configurations where angular momentum is exclusively stored in capillary waves on a deformed cross-section cylinder.
Under these conditions, the calculations allow us to carry out a sensible comparison between the rotational behavior of a  normal fluid
($^3$He) and of an irrotational superfluid  ($^4$He) for fixed values of the atom number and angular momentum per unit length.
In the presence of vortices in addition to
capillary waves, this comparison is obscured as one compares
simply connected configurations for $^3$He cylinders with multiply connected
configurations of vortex-hosting $^4$He cylinders.
Let us recall that in the case of droplets, the presence of vortex arrays dramatically changes the appearance of the droplet;\cite{Pi21}
at fixed angular momentum and atom number in the droplet,
the higher the number of vortices the more compact (i.e., closer to an oblate axisymmetric shape) the droplet becomes.  
Hence, the universal behavior found for classical drops\cite{Bro80} is lost.
At variance, we have found that, independently of  their size, the $^4$He  equilibrium configurations hosting capillary waves alone lay on a
line in the scaled angular momentum and angular velocity plane,\cite{Bro80} disclosing a {\it de facto} nearly universal behavior.
Let us mention that it has recently been found\cite{Ulm23} that, under appropriate experimental conditions,  moderately deformed vortex-free $^4$He drops
prevail when the number of atoms is smaller than about $10^8$.

This work is organized as follows. In Sec. II we  present the methods used
to describe  He cylinders, thoroughly described in Refs. \onlinecite{But11,Anc18,Pi20} in the case of droplets. The results
are discussed in Sec. III, and a summary and discussion are presented in Sec. IV. We outline in  Appendix A  the rationale of how we have defined 
dimensionless angular velocity
and dimensionless angular momentum for the cylinder geometry, and present in Appendices  B  ($^3$He) and C ($^4$He) the results of a simple 
model where the  cross section
of the deformed  cylinder is restricted to be elliptical, treating $^3$He ($^4$He)  as a rotational (irrotational) fluid; we call this model Elliptic Deformations (ED)
model.

\section{Models}

\subsection{Classical approach to viscous systems subjected to centrifugal and surface tension forces}
The incompressible Navier-Stokes equations are solved in a reference frame rotating about a fixed axis perpendicular to the solution domain. 
An arbitrary Lagrange-Euler technique is employed which allows the solution domain to deform and conform to the evolving shape of the drop. 
The rate of displacement of the outer boundary is set by the normal velocity at the outer boundary and surface tension effects are modeled as 
boundary normal stresses that are proportional to the degree of boundary curvature.\cite{walkley2005finite} Models are time dependent and are 
initialized with elliptical domains, with rotation axis passing through the origin, with semimajor and semiminor axes of $(1+\delta)$ and 
$(1+\delta)^{-1}$ where $\delta=0.01$. The small difference from an initial circular shape serves to seed possible non-axisymmetric perturbations. 
At each time step of the simulation, the moment of inertia of the drop is calculated and used to update the rotation rate of the reference frame, 
assuming constant angular momentum. Models are run until the drop shapes achieve a steady state.

The equations are solved using the commercial finite element modeling package Comsol Multiphysics.\cite{comsol} 
Refs.~\onlinecite{But20,But22} give more detailed descriptions of the classical numerical model.

Classical rotating droplets subject to surface tension and centrifugal forces alone are characterized by two dimensionless variables,
angular momentum $\Lambda$ and angular velocity $\Omega$, that allow description of  the sequence of droplet shapes in a universal
phase diagram, independently of the droplet size.\cite{Bro80,Hei06,But11}
For the cylinder geometry, the expressions for $\Omega$ and $\Lambda$  are\cite{But22}  (see also Appendix A)
\begin{eqnarray}
 \Omega &\equiv& \sqrt{\frac{m \, \rho_0 \, R^3}{8 \, \gamma}}\;  \omega
 \nonumber
 \\
\label{eq1}
\\
\Lambda &\equiv&\frac{\hbar}{\sqrt{8 \gamma R^5  m \rho_0}}\, {\cal L}
\nonumber
\end{eqnarray}
where  $\cal{L}$ is the angular momentum per unit length in $\hbar$ units, $\gamma$  and $\rho_0$  are the surface
tension and  atom density of liquid He at zero temperature and pressure, $m$ is the mass of the He atom, and
$R$ is the sharp radius of the  circular He cylinder at rest.
If ${\cal N}$ is the number of He atoms per unit length
of the cylinder,  $R =\sqrt{{\cal N}/(\pi \rho_0)}$.
For liquid $^3$He,  $\gamma$ = 0.1132 K \AA$^{-2}$ and $\rho_0$ = 0.016342 \AA$^{-3}$.
Besides,  $\hbar^2/ m$ = 16.0836 K \AA$^2$.
For liquid $^4$He one has
$\gamma$ = 0.274 K \AA$^{-2}$, $\rho_0$ = 0.021836 \AA$^{-3}$, and
$\hbar^2/ m$ = 12.1194 K \AA$^2$.
Similar to  He droplets made of $N$ atoms, which  are denoted as He$_N$, we shall denote as He$_{\cal N}$
helium cylinders with ${\cal N}$ atoms per unit length.

\subsection{DFT plus semiclassical Thomas-Fermi approach to normal fluid $^3$He}

\begin{figure}[!]
\centerline{\includegraphics[width=1.0\linewidth,clip]{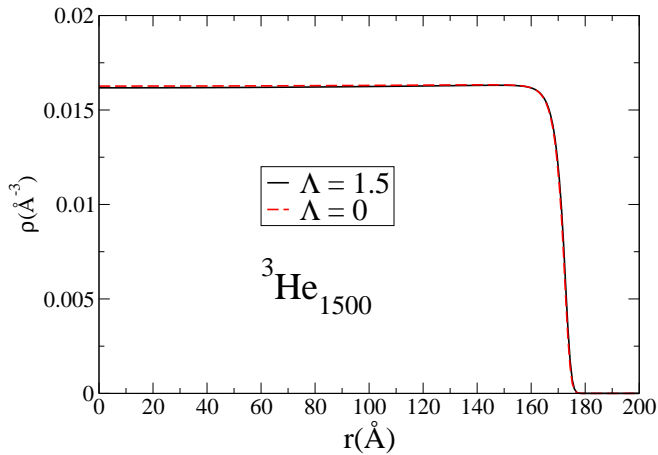}}
\caption{
Density profile of the $^3$He$_{\cal N}$  cylinder with  ${\cal N}=1500$ atoms/\AA{}  for  two circular
configurations corresponding to $\Lambda=0$ (red dashed line) and 1.5 (black solid line). The $\Lambda=1.5$ cylinder  is metastable.
}
\label{fig1}
\end{figure}

We have adapted to the cylindrical geometry the approach used in Ref. \onlinecite{Pi20}
to describe rotating  $^3$He droplets.
Within DFT, the total energy $E$ of a $^3$He$_{\cal N}$ cylinder  at zero temperature is written as a functional
of  the $^3$He  atom density per unit volume $\rho$, here taken from Ref. \onlinecite{Bar97}:
\begin{equation}
E[\rho] =  \int  d {\mathbf r}  \frac{\hbar^2}{2m^*} \tau  +  \int d{\mathbf r} \,{\cal E}_c[\rho]
\equiv \int  d {\mathbf r}  \,{\cal E}[\rho]
\label{eq2}
\end{equation}
The first term is the kinetic energy of $^3$He with an effective mass $m^*$, and $\tau$
 is the kinetic energy density per unit volume, both depending on $\rho$.
 In the  TF  approximation of Ref. \onlinecite{Bar97} (see also Ref. \onlinecite{Str87}),
\begin{equation}
\tau=  \frac{3}{5} (3\pi^2)^{2/3}  \rho^{5/3} +  \frac{1}{18} \frac{(\nabla \rho)^2}{\rho}
\label{eq3}
\end{equation}
The second term in Eq. (\ref{eq3}) is
a Weizs\"acker-type  gradient correction
which is necessary in order to have helium density profiles with an exponential fall-off at the surface.
The energy functional, represented by the energy per unit volume ${\cal E}[\rho]$ in Eq. (\ref{eq2}) within the TF approximation
given by Eq. (\ref{eq3}) accurately reproduces
the equation of state of the bulk liquid and yields the correct value for the $^3$He surface tension.\cite{Bar97}

The equilibrium configuration of the cylinder is obtained by solving the Euler-Lagrange (EL) equation arising from functional
minimization of Eq. (\ref{eq2})
\begin{equation}
\frac{\delta}{\delta \rho}
\left\{\frac{\hbar^2}{2m^*}\tau + {\cal E}_c \right\} = \mu
\label{eq4}
\end{equation}
where $\mu$ is the $^3$He chemical potential corresponding to the number of He atoms per unit length of the cylinder.
Defining $\Psi= \sqrt{\rho}$, Eq. (\ref{eq4}) can be written as a Schr\"odinger-like equation\cite{Bar97}
\begin{equation}
{\cal H}[\rho] \,\Psi  = \mu \Psi
\label{eq5}
\end{equation}
where ${\cal H}$ is the one-body effective Hamiltonian that results from the functional variation.

 \begin{figure}[!]
\centerline{\includegraphics[width=1.0\linewidth,clip]{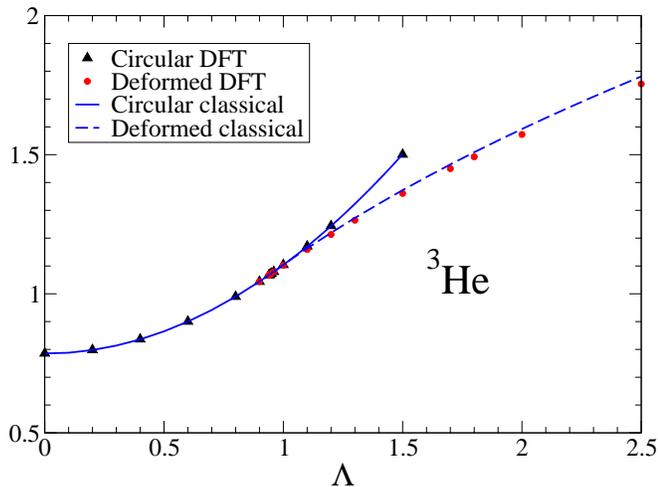}}
\caption{
Reduced DFT Routhian per unit length  $(R[\rho] - \epsilon_0 {\cal N} - E_0)/(8 \gamma R)$ as a function of
rescaled angular momentum.
Black triangles: DFT circular  configurations.  Red circles: DFT deformed configurations.
Solid blue line: classical circular configurations. Dashed blue line: classical deformed configurations.
\label{fig2}
}
\end{figure}

 When the rotating cylinder --made of fermions in the normal phase-- is addressed
in the TF approximation, the  Fermi sphere
is shifted by the motion of the  cylinder as a whole; this adds to its energy density
${\cal E}[\rho]$ a rotational term which has the rigid body appearance\cite{Gra78,Pi20}
\begin{equation}
R[\rho] = \int  d {\mathbf r}  \,{\cal R}[\rho]  =
\int  d {\mathbf r}  \,{\cal E}[\rho]  + \frac{1}{2}  I \omega^2 =  \int  d {\mathbf r}  \,{\cal E}[\rho] + \frac{L^2}{2I}
\label{eq6}
\end{equation}
where  ${\cal R} [\rho]$ is the Routhian density of the cylinder, $L$ is the angular momentum, $\omega$ is the angular
velocity, and $I$ is the moment of inertia.

\begin{figure}[!]
\centerline{\includegraphics[width=1.0\linewidth,clip]{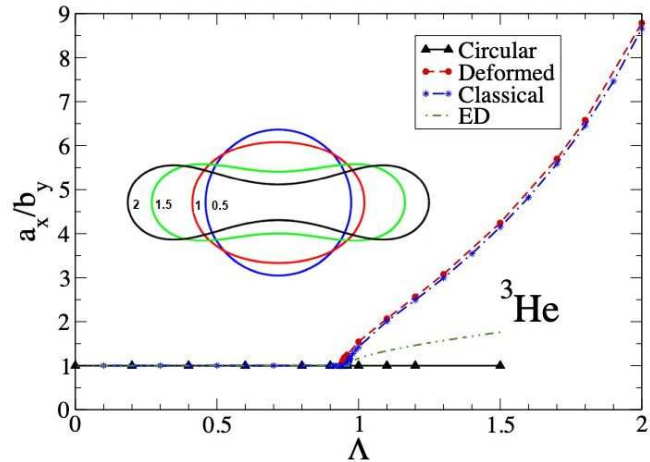}}
\caption{
Aspect-ratio {\it vs.} rescaled angular momentum for the $^3$He cylinder in the DFT, classical and ED approaches.
$AR=1$ corresponds to circular configurations.
The lines are cubic splines of the calculated points.
Also shown are the outlines of classical shapes with angular momenta $\Lambda= 0.5, 1, 1.5$ and 2.
}
\label{fig3}
\end{figure}

Due to the translational invariance of the system along the symmetry axis of the cylinder ($z$ direction), the atom density per unit volume only
depends on the $x$ and $y$ variables
and the integral on $z$ just yields the length $\ell$ of the cylinder. Hence, Eq. (\ref{eq5}) is a two-dimensional partial differential equation
on the $x$ and $y$ variables, and from now on the energy, Routhian and moment of inertia, integrated on the $x$ and $y$ variables
 are quantities per unit length. In particular,
\begin{equation}
I = m \int dx\, dy \,(x^2 + y^2) \rho(x,y)
\label{eq7}
\end{equation}
is the moment of inertia per unit length of the $^3$He cylinder around the $z$-axis, and
$\hbar {\cal L} = I \omega$ is the  angular momentum per unit length.
We stress that the rigid-body moment of inertia is not an imposed ingredient within the DFT-TF framework. It arises naturally from the
TF approximation.\cite{Gra78,Pi20}

\begin{figure}[!]
\centerline{\includegraphics[width=1.0\linewidth,clip]{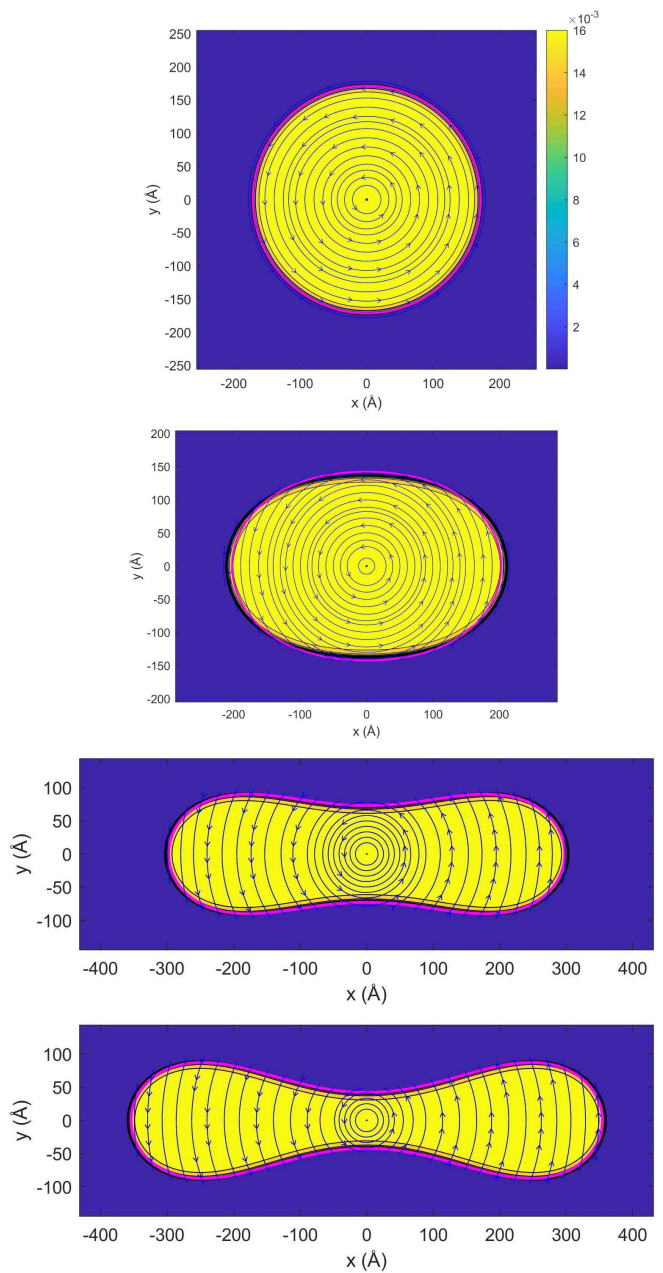}}
\caption{
Two-dimensional densities for the $^3$He$_{1500}$ cylinder in the DFT approach. From top to bottom, they correspond to
$\Lambda=0.5, 1, 1.5$, and 2. The color bar represents the $^3$He density in \AA$^{-3}$. Several streamlines are superimposed.
Also shown are the outlines of classical shapes (magenta lines).
}
\label{fig4}
\end{figure}

Equations (\ref{eq5}) and (\ref{eq8}) have been solved adapting
the $^4$He-DFT-BCN-TLS computing package\cite{Pi17}  to the $^3$He functional.
To take full advantage of the Fast Fourier Transform\cite{FFT} used to carry out the convolution integrals in
the DFT mean field ${\cal H}[\rho]$, we work in Cartesian coordinates and impose periodic boundary conditions (PBC) on the surface of the
box where calculations are carried out. In the $x$ and $y$ directions, this box has to be large enough to accommodate the cylinder in such a way
that the He density is sensibly zero at the box surface,
the effective wave  function $\Psi(\mathbf{r})$ being defined at the nodes of a 2D $N_x \times N_y$ grid spanning the
 $x$ and $y$ directions.  The box is made 3D by adding {\it one single point} in the $z$ direction, $N_z=1$.
 A space step of 0.8 \AA{} has been used.  We have recalculated
several configurations in the circular-to-deformed bifurcation region using a space step of 0.4 \AA{} and have found that the values of the magnitudes 
shown in Table \ref{Table1} are strictly the same.
 The imposed PBC thus make
 $\Psi(\mathbf{r})$ translationally invariant in the $z$ direction as required by the cylinder geometry and in practice one still handles
$N_x \times N_y$ points instead of the $N_x \times N_y \times N_z$ points needed for droplets.
The differential operators  in ${\cal H}[\rho]$  are approximated by 13-point formulas.
The stationary solution corresponding to given values of ${\cal N}$ and ${\cal L}$ is obtained starting from an initial guess
$\Psi_0(\mathbf{r})$ and relaxing it using an imaginary-time step relaxation method.\cite{dft-guide}

It is worth mentioning that at variance with the classical model for viscous liquids subject to centrifugal and surface tension forces,
universality in terms of the scaled $\Lambda$ and $\Omega$ variables  is lost when droplets or cylinders are
described using more refined models that incorporate other effects, e.g., liquid compressibility and surface thickness
effects. Yet, these variables  have been found to be very useful as they allow us to scale the properties of the calculated droplets,
which have a radius of tens of nanometers,\cite{Anc18,Pi20} to those of the experimental ones which have a radius  of hundreds of
nanometers.\cite{Gom14,Ver20} 

For any stationary configuration obtained by solving the EL equation,
a  sharp density surface is determined by calculating
the locus  at which  the helium density equals $\rho_0/2$. Two lengths  are defined corresponding to
 the shortest and largest distances from the $z$ (rotation)
axis to the sharp surface. We call $a_x$ the largest distance, and $b_y$ the shortest one.
 The aspect ratio is defined as $AR=a_x/b_y$, being one for cylinders (circular cross section)
and lager than one otherwise (deformed cross section).

\subsection{DFT approach to superfluid $^4$He}

Within DFT, the energy of the $^4$He$_{\cal N}$ cylinder is written as a functional of the atom density per unit volume $\rho({\mathbf r})$ as\cite{Anc17}
\begin{equation}
E[\rho] = T[\rho] + E_c[\rho] =
\frac{\hbar^2}{2m} \int d {\mathbf r} |\nabla \Psi({\mathbf r})|^2 +  \int d{\mathbf r} \,{\cal E}_c[\rho]
\label{eq9}
\end{equation}
where the first term  is the kinetic energy, with $\rho({\mathbf r})= |\Psi({\mathbf r})|^2$, and
the functional ${\cal E}_c$ contains the interaction term (in the
Hartree approximation) and additional terms which describe non-local correlation effects.\cite{Anc05}

The equilibrium configuration of the cylinder is obtained by solving the EL equation resulting
from the functional minimization of Eq.\ (\ref{eq9}),
\begin{equation}
\left\{-\frac{\hbar^2}{2m} \nabla^2 + \frac{\delta {\cal E}_c}{\delta \rho}  \right\}\Psi({\mathbf r})
 \equiv {\cal H}[\rho] \,\Psi({\mathbf r})  = \mu \Psi({\mathbf r})
\;,
\label{eq10}
\end{equation}
where $\mu$ is the $^4$He chemical potential.

Similarly to the case of $^4$He droplets, to study spinning $^4$He cylinders  we work in the corotating frame at
angular velocity $\omega$,
\begin{equation}
E'[\rho] = E[\rho] - \hbar \omega \, \langle \hat{L}_z \rangle
\label{eq11}
\end{equation}
where $\hat{L}_z$ is the dimensionless angular momentum operator in the $z$-direction;
one looks for solutions of the EL equation resulting from the functional variation of  $E'[\rho]$,
\begin{equation}
\left\{{\cal H}[\rho] \,-\hbar \omega \hat{L}_z\right\} \,\Psi(\mathbf{r})  =  \,\mu \, \Psi(\mathbf{r})
\;.
\label{eq12}
\end{equation}
The differential  operators  in ${\cal H}[\rho]$  and the angular momentum operator are approximated by 13-point formulas.
As in the $^3$He case, the stationary solution corresponding to given values of ${\cal N}$ and ${\cal L}$ is obtained starting from an initial guess
$\Psi_0(\mathbf{r})$ and relaxing it using an imaginary-time step relaxation method.\cite{dft-guide}

 \begin{figure}[!]
\centerline{\includegraphics[width=1.0\linewidth,clip]{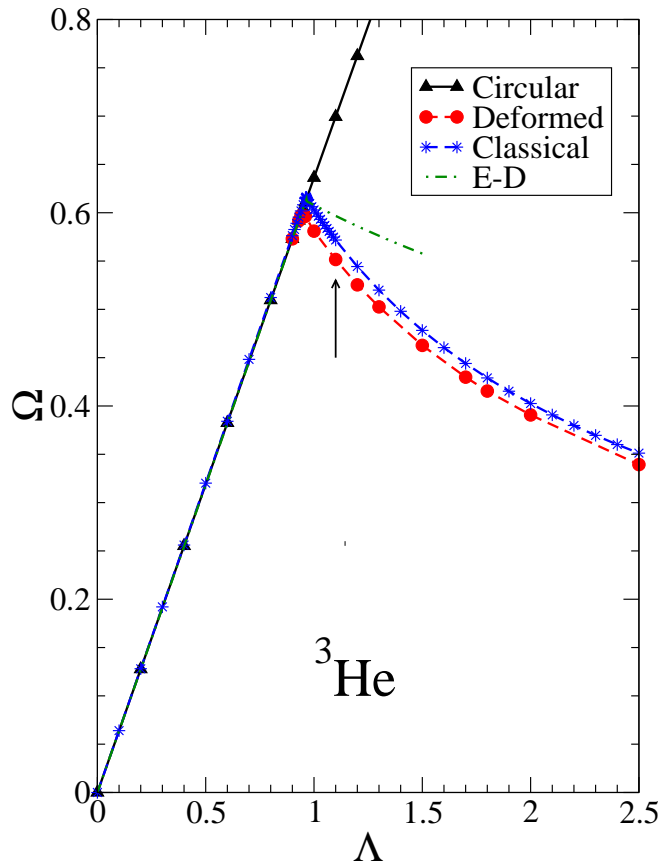}}
\caption{
Rescaled angular velocity $\Omega$ {\it vs.} rescaled angular momentum $\Lambda$ for the $^3$He$_{1500}$ cylinder
in the ED, classical and DFT approaches.
Black triangles, circular  DFT configurations; red circles, deformed DFT configurations; blue asterisks, classical calculations.\cite{But22}
The DFT configurations to the right of the vertical arrow are two-lobed.
The lines are cubic splines of the calculated points.
}
\label{fig5}
\end{figure}

\begin{figure}[!]
\centerline{\includegraphics[width=1.0\linewidth,clip]{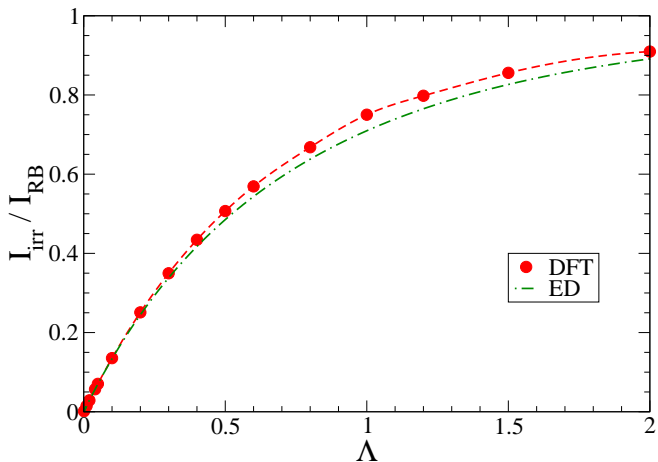}}
\caption{
DFT irrotational moment of inertia $I_{irr}$ in units of the rigid-body moment of inertia $I_{RB}$ for $^4$He cylinders as a function of $\Lambda$.
The lines are cubic splines of the calculated points.
}
\label{fig6}
\end{figure}

Angular momentum can be stored in a superfluid $^4$He sample
in the form of surface capillary waves and/or quantized vortices.\cite{Oco20,Pi21}
Since we are considering only the contribution of capillary waves,
we have used the so-called ``imprinting'' procedure starting the imaginary time relaxation from
 for the  effective wave function
\begin{equation}
\Psi_0(\mathbf{r})=\rho_0^{1/2}(\mathbf{r})\,  e^{i \alpha x y}
\;.
\label{eq13}
\end{equation}
The complex phase $e^{i \alpha x y}$  imprints a surface capillary wave with quadrupolar symmetry
around the $z$ axis,\cite{Cop17} and $\rho_0(\mathbf{r})$  is an arbitrary, vortex-free cylinder
density.  The initial value of $\alpha$ is  guessed, and during the iterative solution of Eq.~(\ref{eq12})
 the shape of the cylinder changes to provide, at convergence, the lowest energy vortex-free configuration
 for the desired ${\cal L}$ value, which requires adjustment of the value of $\omega$ every iteration.

Writing $\Psi(\mathbf{r}) \equiv \phi(\mathbf{r})\, \exp[i\,\cal{S}(\mathbf{r})]$, the velocity field of the superfluid is
\begin{equation}
\mathbf{v}(\mathbf{r}) = \frac{\hbar}{m} {\rm Im}\left\{\frac{\nabla \Psi(\mathbf{r})}{\Psi(\mathbf{r})}\right\}
= \frac{\hbar}{m} \nabla \cal{S}(\mathbf{r})
\label{eq14}
\end{equation}
It can be visualized with streamlines of the superfluid flow.\cite{Pi21} In the $^3$He case, the velocity field consists
of circumferences centered at the rotation axis, with $v(r) = \omega\, r$, being $r$ the distance to the axis.

Equations (\ref{eq10}) and (\ref{eq12}) are two-dimensional partial differential equations depending on the $x$ and $y$ variables
which have been solved  using the $^4$He-DFT-BCN-TLS  computing package.\cite{Pi17}

\section{Results}

\begin{figure}[!]
\centerline{\includegraphics[width=1.0\linewidth,clip]{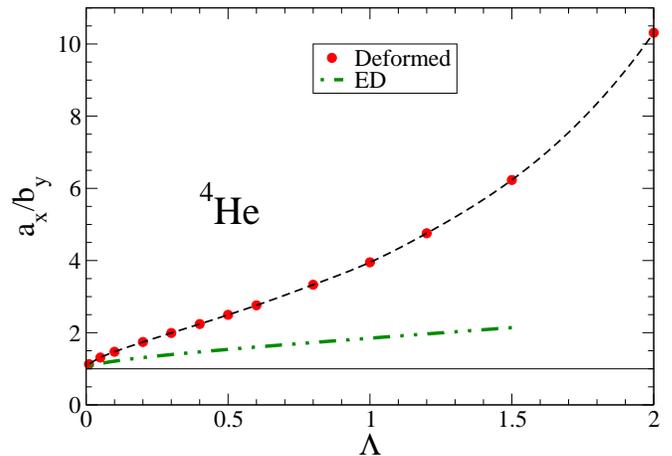}}
\caption{
Aspect-ratio {\it vs.} rescaled angular momentum for the $^4$He cylinder in the ED and DFT approaches.
The lines are cubic splines of the calculated points.
}
\label{fig7}
\end{figure}

We  look for solutions of the EL equation resulting from the functional variation of  $R[\rho]$:
\begin{equation}
\left\{{\cal H}[\rho] \,- \frac{m}{2}\,\left(\frac{L}{I}\right)^2  (x^2+y^2)\right\} \,\Psi(x,y)  =  \,\mu \, \Psi(x,y)  \; .
\label{eq8}
\end{equation}

 We have carried out detailed DFT calculations for a cylinder with ${\cal N} =1500$ atoms/\AA{} which  has a radius $R=170.93$ \AA{} ($^3$He)
 or $R=147.87$  \AA{} ($^4$He) at rest.
As illustrated in Fig. \ref{fig1} for $^3$He, liquid helium is fairly incompressible and hence the cross section area of the rotating He$_{\cal N}$
 cylinder remains sensibly equal to $\pi R^2$ during deformation.

\begin{table}
\begin{tabular}{lccccccc}
\hline
\hline
& $\Lambda$ & $\Omega$ & $a_x$ (\AA)& $b_y$ (\AA) &  $AR$  & $I/I_{cir}$ & $R$ (K/\AA)\\
\hline
C & 0 &0  & 171.41 &  171.41 & 1 &1 &-3641.965 \\
C & 0.20 &0.12768   &171.42  & 171.42        &1 &1.0001& -3639.989 \\
C & 0.40 &0.25525  & 171.45  & 171.45        &1 &1.0006 &-3634.064 \\
C &0.60 &0.38261 &  171.49  & 171.49         &1 &1.0013 &-3624.194 \\
C &0.80 &0.50964   &171.55  & 171.55         &1 &1.0023 &-3610.389\\
C& 0.90 &0.57300  & 171.59  & 171.59         &1 &1.0029 &-3602.013\\
C& 0.93 &0.59198  & 171.60   &171.60         &1 &1.0031 &-3599.310\\
C& 0.94 &0.59830  & 171.61   &171.61         &1 &1.0031 &-3598.389\\
C& 0.95 &0.60463   &171.61  & 171.61  &1 &1.0032 &-3597.458\\
C& 0.96 &0.61095  & 171.62  & 171.62  &1 &1.0033 &-3596.518\\
C& 1.00 &0.63623  & 171.63  & 171.63  &1 &1.0036 &-3592.658 \\
C&1.10 &0.69933  & 171.68  & 171.68   &1 &1.0043 &-3582.326\\
C&1.20 &0.76227 &  171.70  & 171.70   & 1 &1.0051 &-3571.019\\
C& 1.50 &0.95006 &  171.90  & 171.90  &1 &1.0081 &-3531.273\\
\hline
D& 0.90 &0.57297   & 172.43 & 170.78 &   1.010 & 1.0029& -3602.013\\
D& 0.93 &0.59171    &174.20  & 169.05 &  1.030 & 1.0035& -3599.305\\
D& 0.94 &0.59709 & 177.13    & 166.19 &1.066 & 1.0052& -3598.384\\
D &0.95 &0.59563  & 186.72  & 157.02         &1.189 &1.0184 &-3597.461\\
D& 0.96 &0.59633 &  190.93  & 153.11         &1.247 &1.0279 &-3596.539\\
D &1.00 &0.58093  & 210.16  & 136.12         &1.544  &1.0991 &-3592.901\\
D &1.10 &0.55159   &237.00  & 114.55         &2.069 &1.2733 &-3584.134\\
TL& 1.20 &0.52539  & 256.81 &  100.08       &2.566  &1.4583 &-3575.799\\
TL& 1.30 &0.50254  & 273.34  &  88.81        &3.078 &1.6517 &-3567.855\\
TL& 1.50 &0.46269  & 301.70  &  71.08        &4.245  &2.0699 &-3552.930\\
TL& 1.70 &0.42981  & 326.07   & 57.22        &5.698 &2.5254 &-3539.125\\
TL& 1.80 &0.41542  & 337.25   & 51.26        &6.580 &2.7666 &-3532.583\\
TL& 2.00 &0.39071   &357.80   & 40.72        &8.788 &3.2683 &-3520.123\\
TL& 2.50 &0.33938   &403.92    &18.74        &21.550 &4.7035 &-3492.000\\
\hline
\hline
\end{tabular}
\caption{
Configuration characteristics of a  rotating $^3$He$_{\cal N}$ cylinder  with ${\cal N}=1500$ atoms/\AA{} calculated in this work within DFT.
C: circular configurations;
D: deformed, elliptic-like configurations; TL: two-lobed configurations.
$\Lambda$ and $\Omega$ are the dimensionless
angular momentum and velocity, and $R$ is the Routhian per unit length. $AR$ is the aspect ratio  ($AR=1$ for
circular configurations), and  $I/I_{cir}$ is the DFT moment of inertia in units of that of
the  $^3$He$_{1500}$ circular cylinder of sharp radius at $\Lambda=0$,  $I_{cir} = m\,{\cal N}^2/(2\pi \rho_0)$.
\label{Table1}
}
\end{table}

\subsection{$^3$He cylinders}

Table \ref{Table1} collects the DFT results obtained for $^3$He cylinders.
To determine the circular-to-deformed  bifurcation point, one has to compare the
Routhian  $R[\rho]$ of the circular cylinder to that of the deformed cylinder for  the same
$\Lambda$ and ${\cal N}$ values;  the configuration with the smaller $R[\rho]$ is the equilibrium configuration.
 
One can see from Table \ref{Table1}  that the  difference between the Routhians 
 of the circular and deformed cylinders  is very small in a wide interval of $\Lambda$ values between 0.900 and 0.960, 
 which makes rather delicate to determine the bifurcation point within the DFT. 
We take the angular momentum at which the aspect ratio $AR= a_x/b_y$  starts to clearly differ from one as the bifurcation point, 
 having obtained ($\Lambda, \Omega) = (0.90, 0.573)$.   
In the classical model, bifurcation occurs  at ($\Lambda, \Omega) = (0.960, 0.616)$.\cite{But22}

To compare the classical and DFT results for the Routhian (total energy including rotation energy)
per unit length, one has to remember that in classical models only surface  and rotation energy are considered. Consequently,  we have
to identify first the energies that are implicitly involved in the DFT calculation. To this end,
it is convenient to split the energy per unit length of the cylinder $E/\ell$ into different terms, in a way similar in spirit as how the
energy of  the atomic nucleus  is written as a ``mass formula'', namely
\begin{equation}
E/\ell =   \epsilon_0 {\cal N} +  2 \pi R \gamma + E_0 = \epsilon_0 \pi R^2 \rho_0 + 2 \pi R \gamma + E_0
\label{eq15}
\end{equation}
where $\epsilon_0=-2.49$ K is the energy per atom in liquid $^3$He  and $E_0$ is a constant term. Let us mention that
the presence  of a constant term in the nuclear mass formula is common
in the most elaborated ones\cite{Mol87} and it appears after
leading terms of $R^3$ (volume), $R^2$ (surface), and $R$ (curvature) type. In the case of the cylinder, it naturally comes after the surface term.
It is worth noticing that mass formulas  have also been adjusted for $^3$He and $^4$He droplets which include a fairly
large constant term.\cite{Str87,Bar06}

The parameter $E_0$ can be determined from the DFT values at $\Lambda=0$.
For ${\cal N}=1500$ atoms/\AA{},   $R=170.93$ \AA{} and $\epsilon_0 \, {\cal N}$ is   -3735 K/\AA, yielding $E_0= 28.58$  K/\AA{}.

We thus see that the volume and constant energy contributions have to be subtracted from the DFT Routhian for a sensible comparison
with the classical results.
Since in the classical calculations energies per unit length are made dimensionless dividing them by $8 \gamma R$,\cite{But22} the quantity that
can be directly compared with the classical result is the dimensionless reduced DFT Routhian per unit length defined as
\begin{equation}
\frac{1}{8 \gamma R} \,\left\{R[\rho] - \epsilon_0 {\cal N} - E_0\right\}
\label{eq16}
\end{equation}
We have represented it as a function of $\Lambda$ in Fig. \ref{fig2} together with the classical result.
 It can be seen that they agree very well, with some minor differences showing up at large deformations.

Figure \ref{fig3} shows the aspect ratio as a function of the rescaled angular momentum $\Lambda$ for $^3$He cylinders 
obtained with the DFT, classical and ED approaches. The outlines of several classical shapes are drawn in the inset.

We display in Fig. \ref{fig4} the density  of the $^3$He cylinder for several values of $\Lambda$
obtained with the DFT method; superposed to the densities we have plotted several circulation lines.
In the DFT approach the cross section of the cylinder becomes two-lobed at $\Lambda \sim 1.1$.
The outlines of classical shapes are superimposed to the two-dimensional DFT densities.  
It can be seen that they are very similar to the DFT ones except for $\Lambda=1$, for which the DFT density is more deformed because this
configuration is further away from the DFT bifurcation than the classical one is from the classical bifurcation point. This effect 
diminishes at larger deformations where the differences are minor. 

Figure \ref{fig5} shows the $\Omega(\Lambda)$
 equilibrium  line  for $^3$He obtained with the classical  and DFT approaches. This line is very similar for both methods.
 As for $^3$He droplets,\cite{Pi20} the minor differences in the deformed branch are attributed to a better description of the droplet surface
and to quantum kinetic energy contributions in the DFT approach which, together with compressibility effects, are lacking in classical models.
 Also shown in Fig. \ref{fig5}
 is  the result obtained with the ED model as explained in Appendix C. Just away from the bifurcation point the ED approach yields  results very different
 from the two others, indicating that cross section shapes quickly become non-elliptical.
  
\begin{figure}[!]
\centerline{\includegraphics[width=1.0\linewidth,clip]{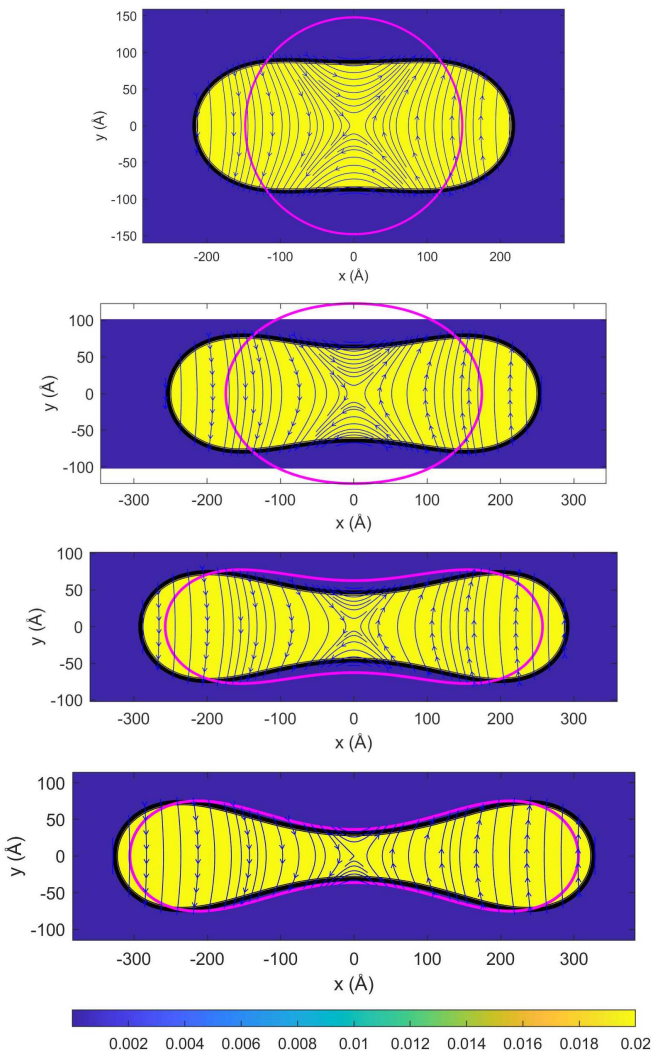}}
\caption{
Two-dimensional densities for the $^4$He$_{1500}$ cylinder in the DFT approach.
 From top to bottom, they correspond to $\Lambda=0.5, 1, 1.5$, and 2.
The color bar represents the $^4$He density in \AA$^{-3}$. Several streamlines are superimposed.
Also shown are the outlines of classical shapes (magenta lines).
}
\label{fig8}
\end{figure}

 \subsection{$^4$He cylinders}

 Table \ref{Table2} collects the DFT results obtained for $^4$He cylinders.
 Since a superfluid system cannot rotate around a symmetry axis,
  the cross section of rotating ($\Lambda \neq 0$) vortex-free $^4$He cylinders must necessarily be non-circular.
  The irrotational moment of inertia, defined as $I_{irr} = \langle L_z \rangle/\omega$, drops to zero as $\Lambda \rightarrow 0$.
  We have plotted $I_{irr}$ in Fig. \ref{fig6}  in units of the rigid-body moment of inertia $I_{RB}$, Eq. (\ref{eq7}).
  It can be seen that $I_{irr}$ approaches $I_{RB}$ at large angular momenta, being rather different even for large deformations.
  For the sake of comparison, we also display the result of the ED model, that works surprisingly well even for $\Lambda$ values for
  which the configuration is no longer elliptic-like but two-lobed.
  The difference between classical and superfluid moments of inertia, also appearing in $^4$He droplets,\cite{Anc15} is a signature of their
  different response to rotations.
  
  \begin{table}
\begin{tabular}{lccccccc}
\hline
\hline
& $\Lambda$ & $\Omega$ & $a_x$ (\AA)& $b_y$ (\AA) &  $AR$  & $I_{irr}/I_{RB}$ & $R$ (K/\AA)\\
\hline
D & 0.001 &0.43503   &150.96   &145.31 & 1.039 &0.0015 &-10466.31\\
D & 0.01 &0.43537   &157.12   &139.25  &1.128 &0.0145 &-10465.04\\
D  &0.02 &0.43561   &160.89   &135.58  &1.187 &0.0289 &-10463.63\\
D & 0.04 &0.43606   &166.24   &130.43  &1.275 &0.0568 &-10460.80\\
D  &0.05 &0.43627   &168.42   &128.35  &1.312 &0.0705 &-10459.39\\
D  &0.10 &0.43772   &177.09   &120.26  &1.473 &0.1355 &-10452.30\\
D  &0.20 &0.44012   &189.77   &108.94  &1.742 &0.2510 &-10438.06\\
D  &0.30 &0.44188   &200.00   &100.37  &1.993 &0.3498 &-10423.77\\
D  &0.40 &0.44295   &209.06    &93.28  &2.241 &0.4344 &-10409.43\\
 TL &0.50 &0.44335   &217.45   & 87.16  &2.495  & 0.5069 &-10395.07\\
 TL &0.60 &0.44304   &225.44    &81.72  &2.759 &0.5691& -10380.70\\
TL  &0.80 &0.44136   &240.38   & 72.21  &3.329 &0.6680 &-10352.24\\
TL  &1.00 &0.43536   &253.38    &64.13  &3.951 &0.7502 &-10323.78\\
TL  &1.20 &0.42614   &269.77    &56.73  &4.755 &0.7980 &-10295.91\\
 TL &1.50 &0.40932   &291.30   & 46.74  &6.233 &0.8558 &-10255.26\\
 TL &2.00 &0.37564   &326.57    &31.68 &10.310 &0.9096 &-10191.55\\
 \hline
 \hline
 \end{tabular}
  \caption{
Configuration characteristics of a  rotating $^4$He$_{\cal N}$ cylinder  with ${\cal N}=1500$ atoms/\AA{} calculated in this work within DFT.
$\Lambda$ and $\Omega$ are the dimensionless
angular momentum and velocity, and $R$ is the Routhian per unit length. $AR$ is the aspect ratio,
and  $I_{irr}/I_{RB}$ is the DFT moment of inertia in units of that of a rigid-body $^4$He cylinder for the same $\Lambda$ value.
D: deformed elliptic-like configurations; TL: two-lobed configurations.
\label{Table2}
}
\end{table}

 Figure \ref{fig7} shows the aspect ratio  as a function of the rescaled angular momentum for $^4$He cylinders obtained with the
DFT and ED approaches.  We display in  Fig. \ref{fig8} the density  of the $^4$He cylinder for several values of $\Lambda$
obtained with the DFT method; superposed to the densities we have plotted several circulation lines.
The cross section of the cylinder becomes two-lobed at $\Lambda \sim 0.5$.
The outlines of classical shapes are superimposed to the two-dimensional DFT densities. 
The difference between classical and DFT densities is apparent up to large $\Lambda$ values, reflecting the distinct rotational response of a normal
fluid from that of a superfluid.

\begin{figure}[!]
\centerline{\includegraphics[width=1.0\linewidth,clip]{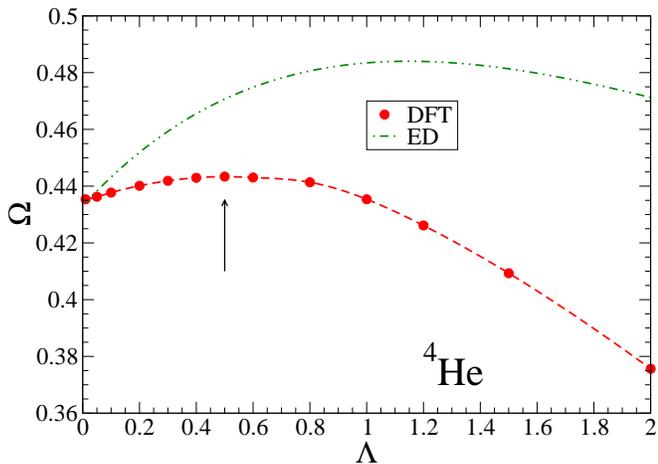}}
\caption{
Rescaled angular velocity $\Omega$ {\it vs.} rescaled angular momentum $\Lambda$ for the $^4$He$_{1500}$ cylinder
in the ED and DFT approaches. The DFT configurations to the right of the vertical arrow are two-lobed.
The lines are cubic splines of the calculated points.
}
\label{fig9}
\end{figure}

 Figure \ref{fig9} shows  the  $\Omega(\Lambda)$ equilibrium  line for $^4$He obtained with the ED and DFT methods. As for $^3$He cylinders, the ED
 approximation quickly becomes inadequate.
 The finite value of $\Omega$ at very small values of $\Lambda$ --also found for $^4$He droplets\cite{Anc18}-- is the equivalent
 of the ``rotational Meissner effect''
 occurring when liquid helium in a rotating cylinder is cooled through the lambda point: at sufficiently slow rotational speeds the superfluid
 forms in a state of zero total angular momentum, causing the container to rotate faster.\cite{Hes67}
 
\begin{figure}[!]
\centerline{\includegraphics[width=1.0\linewidth,clip]{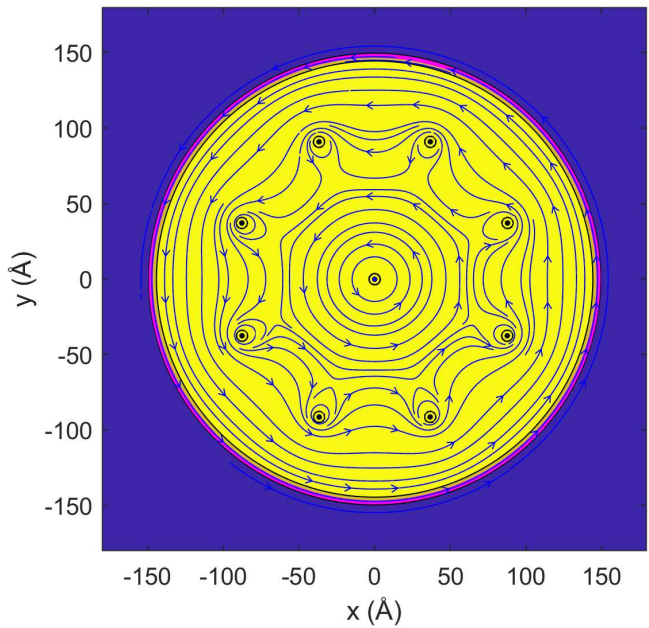}}
\caption{
Two-dimensional density for the $^4$He$_{1500}$ cylinder hosting nine vortex lines at  $\Lambda=0.5$ in the DFT approach. 
Several streamlines are superimposed.
Also shown is the outline of classical shape (magenta line).
}
\label{fig10}
\end{figure}

\section{Discussion and outlook}

As for helium droplets,\cite{Pi20} we have expectedly found that the rotating
behavior of normal fluid $^3$He cylinders is very similar to rotating incompressible, viscous
cylinders only subject to surface tension and centrifugal forces.
Even for fine details such as the aspect ratio as a function of rescaled angular momentum or
the $\Omega(\Lambda)$ equilibrium line, 
we have found a good agreement between classical and DFT  results.

Figures \ref{fig4} and \ref{fig8} clearly illustrate how different is instead
the response to rotation of the normal fluid  $^3$He cylinder from that of superfluid $^4$He, especially at
moderate angular momentum values, i.e., at aspect ratios not very different from one.
Only when $\Lambda$ is large do the density profiles become similar. It is worth recalling that the number of droplets having large
deformations is found to be very small in the experiments.\cite{Lan18} It is also worth seeing that even for these very deformed cylinders,
the moment of inertia of superfluid $^4$He cylinders is 10-20\% smaller that the rigid-body value (Fig. \ref{fig6}).

A close look at the appearance of the streamlines in Figs. \ref{fig4} and \ref{fig8} shows that while $^3$He cylinders do rotate and
streamlines are circumferences around the rotation axis,  $^4$He cylinders do not; the streamlines
do not correspond to a rigid rotation, but to an irrotational flow.
Lets us remember that the fluid motion is a combination of translation, rotation and deformation of the fluid elements, and only when vorticity (defined as
$\nabla \times \mathbf{v}$)\cite{Guy15} is nonzero, may one speak of a true rotation. Vorticity  is distributed inside
the cylinder in the normal phase $^3$He and it equals $2 \omega$ as for a rotating rigid body in steady rotation.
Since the superfluid flow is irrotational, $\nabla \times \mathbf{v}=0$ for $^4$He.
In this case, fluid elements translate and deform, but do not rotate. An illuminating
discussion of this behavior, based on a  paper by  Tritton,\cite{Tri82} can be found in Ref. \onlinecite{Don91}.
The different behavior of $^3$He and vortex-free $^4$He cylinders in rotation is also apparent in the $\Omega(\Lambda)$ equilibrium line and in the
moment of inertia as a function of the angular momentum, Figs. \ref{fig5}, \ref{fig6} and \ref{fig9}. 

When the superfluid $^4$He sample hosts linear vortices,
the situation dramatically changes, as the vortex array tends to confer to the droplet or cylinder the appearance of a rotating rigid body.\cite{Anc18,Osb50}
To exemplify this key issue, we have redone the calculation of the rotating $^4$He cylinder at $\Lambda=0.5$ when it hosts a large vortex array. 

One single vortex  along the axis of the circular cylinder has an angular momentum per unit length $\cal{L}= \cal {N}$.\cite{Pit16} From the
definition of $\Lambda$, Eq. (\ref{eq1}), it corresponds to a rather small value if $\cal{N}$=1500 atoms/\AA{}, namely $\Lambda=0.0898$.
We have imprinted a nine vortex array  to the cylinder by using the wave function\cite{Pi21}
\begin{equation}
\Psi_0(\mathbf{r})=\rho_0^{1/2}(\mathbf{r})\,  
\prod _{j=1}^{n_v} {(x-x_j)+i (y-y_j) \over \sqrt{(x-x_j)^2+(y-y_j)^2}}
\label{eq17}
\end{equation}
with $n_v=9$, where $(x_j, y_j)$ is the initial position of the $j$-vortex core. We have started from a deformed cylinder 
and have chosen an initial vortex array with a centered vortex and eight vortices around it at the same distance $d$, which is the
lowest energy configuration in the classical case.\cite{Cam79} During the iterative solution of Eq.~(\ref{eq12})
both the vortex core structure and positions, and the  shape of the cylinder may change
to provide at convergence the lowest Routhian configuration.

The effect caused by the presence of the vortex array on the morphology of the free-standing cylinder can be appreciated in Fig. \ref{fig10}:
the very deformed vortex-free cylinder, with $AR=2.50$ (see Table \ref{Table2}), has become circular ($AR=1.001$), coinciding
with the classical result. 
 For smaller $n_v$ values the cylinder may still be deformed, but 
not as much as the vortex-free cylinder. For instance, when $n_v=6$, $AR=1.25$. 
In this case, vortices and capillary waves coexist.\cite{Oco20,Pi21}  
We have found that for the $\Lambda=0.5$ value used here, the $n_v=9$ configuration is more stable than the vortex-free one.

It is also worth noticing that for a $n_v$ vortex array in a rotating cylinder of radius $R$, the total angular momentum can be written as\cite{Hes67b}
\begin{equation}
{\cal L}= {\cal N} \sum_{j=1}^{n_v} \left[1-\left(\frac{d_j}{R} \right)^2\right] 
\label{eq18}
\end{equation}
where $d_j$ is the distance of the $j$-vortex to the symmetry axis, 
which in our case reduces to ${\cal L}= {\cal N} + 8\,{\cal N} [1- (d/R)^2]$. This expression yields $d=96.8$ \AA{} for $R=147.87$ \AA{}
and $\Lambda=0.5$, in perfect agreement with the averaged DFT result.

All configurations obtained in this work, in particular those displayed in Figs. \ref{fig4} and \ref{fig8}, are stationary in the co-rotating frame --the framework that
rotates with angular velocity $\omega$ with respect to the laboratory frame. Consequently, they would be seen from the laboratory as if they
were rotating like a rigid body with the angular frequency $\omega$ imposed to obtain them.\cite{Sei94} The  $^3$He cylinder would undergo a
true rotation, but this rotation would only be apparent for the $^4$He cylinder. Examples of apparent rotations in the case of $^4$He droplets, obtained
within time-dependent DFT, can be found in Refs. \onlinecite{Cop17,Pi21}.

 Ongoing experiments on mixed helium droplets,\cite{Oco21}
which exhibit a core-shell structure with a crust made of $^3$He atoms in the normal state and a superfluid
core mostly made of $^4$He atoms, and calculations on mixed droplets made
 of immiscible  viscous fluids\cite{But22} call for extending the DFT calculations carried out for mixed helium droplets\cite{Pi20} to larger sizes, also relaxing the
 constraint  that $^3$He and $^4$He moieties are concentric. When the $^4$He core displaces with respect to the center of mass of the
droplet,  the moment of inertia increases from that of centered drops,  influencing  how angular
  momentum is stored in the mixed droplet. In particular, it might affect quantum vortex nucleation in the $^4$He core, which could be hindered.
 The cross section of deformable cylinders is a reasonable representation of the density of a large rotating helium droplet on the plane of symmetry
perpendicular to the rotation axis.\cite{Oco20}  The use of cylindrical geometry would allow the extension of DFT calculations to helium drops of larger cross section
that would otherwise be computationally prohibitive.

\begin{acknowledgments}
This work has been  performed under Grant No.  PID2020-114626GB-I00 from the MICIN/AEI/10.13039/501100011033
and benefitted from COST Action CA21101 ``Confined molecular systems: form a new generation of materials to the stars'' (COSY) 
supported by COST (European Cooperation in Science and Technology).
J.~M.~E. acknowledges support from the Spanish Research Agency MCIN/AEI/10.13039/501100011033
through the Severo Ochoa Centres of Excellence programme (grant SEV-2017-0706).
\end{acknowledgments}

\appendix
\setcounter{equation}{0}
\section{Rotating $^3$He cylinders held together by surface tension: the general case}

In this Appendix we outline how dimensionless angular momentum $\Lambda$ and
dimensionless angular velocity $\Omega$ can be introduced in the case
of an incompressible  cylinder of length $\ell$ and radius $R$, modeled by viscous cylinders
subject to surface tension and centrifugal forces alone. To connect with the 2D DFT results, the length $\ell$ will be taken as infinite, and
eventually all extensive quantities will be referred to per unit length.
 We closely follow the procedure by Brown and Scriven in the case of droplets.\cite{Bro80}

 In cylindrical coordinates $(r,\phi,z)$, the   radial vector to the surface
 is described by $\mathbf{r} = R f(\phi) \mathbf{\hat{r}}$, where
 $\mathbf{\hat{r}}$ is the unit vector in the radial direction and $\phi$ is the azimuthal angle.
 This representation handles circular, elliptic-like and multi-lobed cylinders, with the limitation that $\mathbf{r}$ must
 not intersect the surface of the cylinder more than once.

 If the cylinder rotates around its axis ($z$-axis) at  angular velocity $\omega$, the energy per unit length is
 given by
 \begin{equation}
 E =  \gamma L_c + \frac{1}{2} I \omega^2 \; ,
 \label{A1}
 \end{equation}
where $\gamma$ is the surface tension, $L_c$ is the perimeter of the cross section of the cylinder
\begin{equation}
 L_c =  2 R \int^{\pi}_0 d \phi \, \sqrt{f^2(\phi) + \left(\frac{\partial f}{\partial \phi}\right)^2} \; ,
 \label{A2}
 \end{equation}
and $I$ is the moment of inertia per unit length
\begin{equation}
 I =  \frac{1}{2} m \rho_0 R^4 \int^{\pi}_0 d \phi \, f^4(\phi) \equiv  m \rho_0 \, R^4 \, {\cal I}
 \label{A3}
 \end{equation}
where  ${\cal I}$ is the dimensionless moment of inertia per unit length.
Writing the energy per unit length in units of $2 \gamma R$, we have
\begin{equation}
\frac{E}{2 \gamma R} = \bigintsss_0^{\pi} d \phi \, \left\{\sqrt{f^2(\phi) + \left(\frac{\partial f}{\partial \phi}\right)^2}
+   \frac{m \rho_0 \omega^2 R^3}{8 \gamma}   \, f^4(\phi)\right\}
\label{A4}
\end{equation}
The ratio
\begin{equation}
\Sigma \equiv \Omega^2 = \frac{m \rho_0 \,\omega^2\,R^3}{8 \gamma}
\label{A5}
\end{equation}
is called rotational Bond number,\cite{Bro80,Ben91,Cha65} and is the dimensionless measure of the square of angular velocity $\Omega$.
A dimensionless angular momentum $\Lambda$ is introduced such that $\Lambda= {\cal I}\, \Omega$. This yields Eqs. (\ref{eq1}) in the
main text.
Eq. (\ref{A4}) shows that within this model the solution is universal and can be obtained once for all rotating cylinders.

We want to comment that our definition of $\Omega$ coincides with that of Ref. \onlinecite{Ben91}; unfortunately, the definition of $\Lambda$ is not given
in that reference.

\section{Rotating $^3$He cylinders held together by surface tension: elliptic deformations}

It is illustrative to address the classical rotating cylinder when deformations are restricted to be elliptic, as it is nearly analytical and
it is expected to be a fair approximation to describe the circular to deformed bifurcation.
Proceeding as in previous Appendix A, the surface energy per unit length of the cylinder
is written as $E_s = \gamma L_c$, where
\begin{equation}
L_c = 4 a \int^{\pi/2}_ 0 d \phi \sqrt{ 1- e^2 \sin^2\phi} \equiv 4 a\, \mathbf{E}(e)
\label{B1}
\end{equation}
is the perimeter of the ellipse
\begin{equation}
\frac{x^2}{a^2} + \frac{y^2}{b^2} = 1
\label{B2}
\end{equation}
with eccentricity $e = \sqrt{a^2-b^2}/a$  (we take $a \geq b$).
If the fluid is incompressible, $\pi a b = \pi R^2$. In Eq. (\ref{B1}), $\mathbf{E}(e)$ is the complete elliptic integral
of the second kind\cite{Gra07}
\begin{equation}
\mathbf{E}(e) =  \int^{\pi/2}_ 0 d \phi \, \sqrt{ 1- e^2 \sin^2\phi}
\label{B3}
\end{equation}
Defining $\xi= a/R= (1-e^2)^{-1/4}$, where $R$ is the radius of the cylinder at rest,
the moment of inertia per unit length can be expressed as
\begin{equation}
I= \frac{\pi}{4} m \rho_0 \, R^4 \,\frac{\xi^4+1}{\xi^2}
\label{B4}
\end{equation}
The energy per unit length  in units of $8 \gamma R$ is
\begin{equation}
\frac{E}{8 \gamma R} = \frac{1}{2} \xi \,\mathbf{E}(e) +\frac{2}{\pi} \Lambda^2 \frac{\xi^2}{\xi^4+1}
\label{B5}
\end{equation}
which for the circular cylinder reduces to
\begin{equation}
\frac{E}{8 \gamma R} = \frac{\pi}{4}  + \frac{1}{\pi} \Lambda^2
\label{B6}
\end{equation}

Determining the equilibrium configuration for a given $\Lambda$ amounts to solving for $e$ (or $\xi$) the algebraic equation
$d{\cal E}/d\xi=0$:
\begin{equation}
\mathbf{E}(e) + \frac{2}{\xi^4 - 1} [\mathbf{E}(e) - \mathbf{K}(e)] + \frac{8}{\pi} \Lambda^2 \, \frac{\xi (1- \xi^4)}{(\xi^4+1)^2} = 0
\label{B7}
\end{equation}
where $\mathbf{K}(e)$ is the complete elliptic integral of the first kind\cite{Gra07}
\begin{equation}
\mathbf{K}(e) =  \int^{\pi/2}_ 0 d \phi \, \frac{1}{\sqrt{ 1- e^2 \sin^2\phi}}
\label{B8}
\end{equation}
The determination of the equilibrium configuration is facilitated by the existence of accurate easy-to-use approximations for
$\mathbf{K}(e)$ and $\mathbf{E}(e)$.\cite{Abr}
The dimensionless angular velocity is obtained from the $\xi$ value of the equilibrium configuration at the given $\Lambda$ as
\begin{equation}
\Omega =  \frac{4}{\pi} \frac{\xi^2}{\xi^4+1}\, \Lambda
\label{B9}
\end{equation}
which reduces to $\Omega = 2 \Lambda/\pi$ for the circular cylinder.

We have found that the circular-to-elliptical shape transition occurs at $\Lambda =0.966$, i.e., at
$(\Lambda,\Omega)=(0.966,0.612$).
Notice that $\Sigma$ at the bifurcation point is $\Sigma=3/8$,\cite{Ben91} which yields  $\Omega=\sqrt{3/8}=0.612$.
 At bifurcation,  $\Omega = 2 \Lambda/\pi$ and one has
$\Lambda=0.966$ instead of the value close to 2 shown in Fig. 4a of Ref. \onlinecite{Ben91} (see also Ref. \onlinecite{But22}).
For this reason, we have inferred that the definition
of $\Lambda$ in that reference likely is  a factor of two larger than ours.  As shown in Fig. \ref{fig5},
the elliptic deformations model is unrealistic for $\Lambda \gtrsim 1.1$ and misses the appearance of two-lobed configurations.
We conclude that  deformations after bifurcation quickly become complex and representing the cross section of the
deformed cylinder by an ellipse  is a rough approximation.

\section{Rotating $^4$He cylinders held together by surface tension: elliptic deformations}

It is also illustrative to consider the case of a vortex-free superfluid $^4$He elliptic cylinder in which angular momentum is stored only in capillary waves.
The flow is irrotational and its velocity  derives from a  velocity potential
 $\chi(x,y)= \alpha xy$,   which  yields\cite{Pit16,Boh75,Cop17}
\begin{equation}
{\mathbf v} = \nabla \chi(x,y) = \omega \, \frac{a^2 -b^2}{a^2+b^2} \, (y,x,0)
\label{C1}
\end{equation}
The $z$-component of
\begin{equation}
m \rho_0 \int dx \,dy \, {\mathbf r} \times {\mathbf v}
\label{C2}
\end{equation}
is the angular momentum per unit length
\begin{equation}
{\cal L}=  \frac{\pi}{4} m \rho_0  \, \frac{(a^2-b^2)^2}{a^2+b^2}\, a b \,\omega \equiv I_{irr} \, \omega
\label{C3}
\end{equation}
where $I_{irr}$ is the irrotational moment of inertia per unit length. Writing it in terms of $R$ and $\xi$
we have
\begin{equation}
I_{irr}=  \frac{\pi}{4} m \rho_0  \, R^4 \frac{(\xi^4-1)^2}{\xi^2 (\xi^4+1)}
\label{C4}
\end{equation}

Notice that the ratio between irrotational (Eq. (\ref{C4})) and rotational, rigid-body (Eq. (\ref{B4})) moments of inertia is
\begin{equation}
\frac{I_{irr}}{I_{RB}} =  \left(\frac{\xi^4-1}{\xi^4+1}\right)^2 \,
\label{C5}
\end{equation}
which is zero for a circular cylinder ($\xi=1$). The energy per unit length  in units of $8 \gamma R$ is
\begin{equation}
\frac{E}{8 \gamma R} = \frac{1}{2} \xi \,\mathbf{E}(e) +\frac{2}{\pi} \Lambda^2 \, \frac{\xi^2 (\xi^4+1)}{(\xi^4-1)^2}
\label{C6}
\end{equation}

As for the rotational $^3$He fluid, determining the equilibrium configuration for a given $\Lambda$ amounts to solving for $e$
(or $\xi$) the algebraic equation $d{\cal E}/d\xi=0$:
\begin{eqnarray}
&& \mathbf{E}(e) + \frac{2}{\xi^4 - 1} [\mathbf{E}(e) - \mathbf{K}(e)]
\\
&&
\nonumber
- \frac{8}{\pi} \Lambda^2 \, \frac{\xi}{(\xi^4-1)^3} \,[\xi^8 + 6\xi^4+1] = 0
\label{C7}
\end{eqnarray}
The dimensionless angular velocity $\Omega$ is obtained from the $\xi$ value of the equilibrium configuration at the
given $\Lambda$ as
\begin{equation}
\Omega =\frac{4}{\pi}  \, \frac{\xi^2 (\xi^4+1)}{(\xi^4-1)^2} \, \Lambda
\label{C8}
\end{equation}

\end{document}